# Classification with Quantum Machine Learning: A Survey

Zainab Abohashima [1] | Mohamed Elhoseny [2] | Essam H. Houssein [3] | Waleed M. Mohamed [3]

[1] Faculty of Computer Science, Nahda University, Egypt
[2] Faculty of Computers and Information, Mansoura University, Egypt
[3] Faculty of Computers and Information, Minia University, Egypt

**Abstract**
**Due to the superiority and noteworthy progress of Quantum Computing (QC) in a lot of applications such as cryptography, chemistry, Big data, machine learning, optimization, Internet of Things (IoT), Blockchain, communication, and many more. Fully towards to combine classical machine learning (ML) with Quantum Information Processing (QIP) to build a new field in the quantum world is called Quantum Machine Learning (QML) to solve and improve problems that displayed in classical machine learning (e.g. time and energy consumption, kernel estimation). The aim of this paper presents and summarizes a comprehensive survey of the state-of-the-art advances in Quantum Machine Learning (QML). Especially, recent QML classification works. Also, we cover about 30 publications that are published lately in Quantum Machine Learning (QML). we propose a classification scheme in the quantum world and discuss encoding methods for mapping classical data to quantum data. Then, we provide quantum subroutines and some methods of Quantum Computing (QC) in improving performance and speed up of classical Machine Learning (ML). And also some of QML applications in various fields, challenges, and future vision will be presented.**

## 1 | INTRODUCTION

As is well known the role of machine learning (1-4) in data analysis, feature selection, making decision, pattern classification and future predictions of various applications with achieving better accuracy and performance without human decisions. but an enormous increase in types of data (i.e., images, text, videos, and record audios) and current computational resources appeared problems of machine learning such as high-cost learning and kernel estimation. Over the last three decades, we noticed using quantum computing (QC) in various applications as cryptography (5), artificial intelligence (AI) (6), and communication (7). As a result of the superiority and advancement of quantum computing for solving many problems as factorization by Shor's algorithm (8), and search in an unstructured database by Grover's algorithm (9).

Quantum computing based on postulates and characteristics of quantum mechanics (i.e., quantum bits (Qubits), interference, superposition, and entanglement) to information processing. qubit can be one state, zero state, or a combination of two states at the same time known as linear superposition, unlike classical bit that can represent one value either 0 or 1 to store information (7). Qubit state is a unit vector in Hilbert space. mathematically to represent qubit state, we use ket- notation, qubit in state zero is $|0\rangle = [1 \quad 0]^T$ and qubit in state one is $|1\rangle = [0 \quad 1]^T$. a qubit is represented as a linear superposition of both basis states simultaneously:

$$|\psi\rangle = \begin{pmatrix} \alpha \\ \beta \end{pmatrix} = \alpha|0\rangle + \beta|1\rangle \qquad (1)$$

Where the coefficients $\alpha$ and $\beta$ are probability amplitudes may be complex numbers and $|\alpha^2| + |\beta^2| = 1$.

Recently, different algorithms and approaches have been published in machine learning (ML) depends on quantum computing laws to introduce the new concept called quantum machine learning (QML) (4) (10, 11) with goal improvement of classical ML (12). QML appeared in more than a paradigm. the first paradigm, quantum ML, in 2014 Rebentrost, p et al (24) presented a quantum version of support vector machine (QSVM) for classification of big data. QSVM achieved logarithmic speed up over classical counterparts. The second paradigm, Quantum inspired ML, in 2019 Sergioli et al. (28) proposed a binary quantum classifier inspired by the formalism of quantum theory and this classifier achieved a higher performance compared with different classical models. In 2019 Havlicek et al. (40) presented a quantum-classical classifier depends on the concept of the variational quantum circuit and this paradigm named hybrid quantum-classical ML. Also, various models have developed in Quantum Neural Networks (QNNs) such as QNNs (25), Quantum inspired NNs (37), Quantum multi perceptron NNs, quantum dot NNs, and quantum convolutional NNs. in detail, all these paradigms will be presented and comparison among them in section 2. QML is used in several applications as medical data classification (30,43), recommendation systems, and big data processing (24).

Quantum subroutines are the lifeblood of quantum and QML algorithms. Quantum subroutines have a major role in speed up and improve the performance of algorithms such as HHL algorithms (88) for solving "linear system equations" in exponential speed up based on quantum phase estimation (QPE) and quantum matrix- inversion subroutines and Grover's algorithm based on amplitude amplification. Other subroutines as sampling, quantum annealing, and quantum fourier transform (QFT).

The main objectives of this survey introduce a classification scheme with QML. To implement a classical algorithm on a quantum computer or in general any quantum algorithm consists of the main three phases (13, 14): encoding, quantum computation, and decoding phase. first of all, the encoding phase that means mapping data from classical to quantum states. In the second step, quantum computation depends on type QML algorithms. the last step, decoding phase. In addition to illustrating the major phases of quantum classification problems to answer the question of survey what is the classification scheme with quantum machine learning?

We will open a discussion about two questions in QML algorithms. Why do we proceed towards Quantum Machine learning (QML)? How does the concept of quantum computing enhance classical machine learning? The main objectives of this survey are:

1. Summarize and organize the most recent research works to pave the way for researchers in quantum machine learning.
2. Analyze techniques and show the best and most used methods in classification real problems.
3. Provide the readers with various quantum methods to enhance classical ML and some of the quantum subroutines.
4. Introduce quantum classification scheme in the quantum world.
5. Present some of the challenges, future directions, and applications of quantum machine learning

The remainder of the survey can be divided into six sections: section 2 presents the major and recent approaches to quantum machine learning techniques especially quantum classification algorithms classified into 3 categories QML, quantum-inspired ML, and hybrid quantum-classical ML. Also, summarizes these techniques in table 1. Next, we present the concept of classification with QML, two methods of encoding data, variational quantum circuits, and, proposed quantum classification scheme in section 3. Thereafter, a discussion of the survey questions in section 4. section 5 introduces some of the challenges and future directions of QML. We provide many applications of QML in section 6 Finally, concludes the survey in section 7

## 2 | LITERATURE REVIEW

Here, we present exhaustive previous work about QML. As well as, we classify quantum machine learning algorithms into three approaches: QML, quantum-inspired ML, and Hybrid quantum-classical ML and summarized in table 1 & table 2. The classification of approaches based on information processing devices (a type of algorithm) is classical or quantum and type of data are classical or quantum (15, 16).

Montanaro, A. (17) introduced an overview of quantum algorithms and some of the applications in different areas. Jeswal, S. et al. (18) presented various quantum neural network techniques and their applications in real-world problems. the authors discussed each technique in detail and noted that QNNs are more powerful and enhance computational efficiency in contrast to classical NNs. Benedetti, M et al. (19) offered an overview of hybrid quantum-classical models based on parameterized quantum circuits and the application of hybrid systems in supervised and generative learning. They also provided a framework for components of models as an example, variational circuit and encoder circuit. In (20), the authors given an overview of the progress of quantum computing and applications. Also, the authors discussed various quantum technologies for quantum computers scalability such as error correction and going forward. Ciliberto, C et al.(21) presented a review about QML and challenges. AS well, the authors provided some quantum subroutines especially quantum linear algebra and how a quantum computer works with data. Besides, this review illustrates Quantum neural networks.

*A. Quantum machine learning algorithms*

The first approach, quantum machine learning algorithms are quantum versions from conventional ML. as well, algorithms that can be executed on the real quantum device. Dennis, et al. (22) implemented SVM on quantum annealer device (23) (DW2000Q) called QA-SVM. The authors used quantum annealer to train and optimize SVM depends on QUBO equation to minimize cost energy. As well, the authors utilized some of feature of quantum annealing (i.e. reverse annealing, and special annealing schedules to improve final results. Rebentrost, P. et al (24) presented SVM algorithm runs quantum computer on and depends on a non-sparse matrix called QSVM. QSVM is a big data binary classifier. As well, it works with a large number of features and samples in complexity logarithmic. da Silva, A. et al (25) introduced a new quantum neural network named "quantum perceptron over a field" (QPF) and its learning algorithm (SAL). the learning algorithm (SAL) based on superposition feature and quantum operator. Also, it performs NN architecture with polynomial time. QPF overcomes on limitations of quantum perceptron models. In (26) the authors proposed a version of linear regression called quantum linear regression.it works on quantum data with N- dimensions of features in logarithmic time.

*B. Quantum-inspired machine learning*

The second approach, quantum-inspired machine learning that applies the principles of Quantum Computing (QC) to improve classical methods of machine learning (ML).

Prayag et al. (27) introduced a new quantum-inspired binary classifier (QIBC), the basic idea is based on decision theory, classical ML and theory of quantum detection that utilize one of the laws of quantum mechanics, superposition to increase the higher degree of freedom in decision making. The proposed classifier can be achieved high precision, recall and F-measure comparable with KNN, and SVM and other classical techniques. Sergioli et al. (28) proposed a novel quantum-inspired classifier for binary supervised learning called Helstrom Quantum Centroid (HQC) based on density matrices and formalism of quantum theory. The authors evaluated the performance of their model by fourteen datasets compared to different classical models. Ding et al. (29) proposed a novel algorithm inspired by the quantum support vector machines(SVM) to solve classification problems in exponential speedup. The main idea for the algorithm based on linear transformation.

Sergioli et al. (30) introduced a Quantum Nearest Mean Classifier (QNMC) based on the idea of classical minimum distance classifier. The algorithm consists of three main steps: firstly, density pattern (Encoding) to transform each classical data point to a quantum object. Secondly, Quantum centroid to calculate a distance among density patterns in order to classify unknown quantum objects to the right class. lastly, decoding to transform the final classification result into the classified data. The algorithm achieved higher accuracy in many medical data sets than the classical counterpart (NMC) exclusively cancer data set. Dang, Y et al. (31) proposed a new model based on Quantum KNN and parallel computing for image classification. their model improved efficiency and classification performance. Chen, H. et al. (32) used powerful parallel computing to introduce an inspired Quantum K Nearest-Neighbor (QKNN) based on one of the well-known properties of QC is superposition to obtain parallel computing and "quantum minimum search algorithm" to achieve speed up the search.

Lu, S et al (33) proposed a quantum version for decision tree classifier. The quantum model depends on quantum entropy impurity and quantum fidelity measure. In (34) the authors developed a new model Quantum Support Vector Clustering for big data depended on Quantum SVM, quantum Gaussian kernel and quantum polynomial kernel. In (35), The authors proposed new framework quantum clustering based on the Schrödinger equation. Yu, C.et al. (36) presented improved quantum techniques for ridge regression. This algorithm based on two subroutines quantum K-fold cross-validation and quantum state encoding. In (37) the authors introduced a novel Quantum inspired Neural network. they named Autonomous Perceptron Model (APM) depends on one from quantum mechanics characteristics a quantum bit. The proposed model accomplished higher accuracy with less complexity time comparable to other tradition-al algorithms.

*C. Hybrid quantum-classical machine learning*

The last approach, hybrid quantum-classical machine learning are the algorithms that combine quantum algorithms and classical (traditional algorithms) to obtain higher performance and decrease in the learning cost.

Soumik et al. (38) used the quantum circuit to present a new variational quantum classifier with a single quantum system (Qu *N* it) to encode data in N-dimensional with a training algorithm called "single-shot training". the key advantage of single-shot training uses a fewer parameter for training and achieve a higher precision. In (39), the authors introduced a new quantum algorithms based on many subroutines as a quantum oracle, counting, amplitude amplification, and quantum amplitude estimation for feature selection named (HQFSA) with purpose enhancement of performance ML techniques. The proposed algorithm accomplished quadratic time complexity and better performance in some of the cases. The main disadvantage of HQFSA is running on a quantum simulator only.

Havlicek et al. (40) suggested two different models of quantum support vector machines. the first, the variational quantum SVM based on quantum variation circuit. The variational quantum SVM classifier requires two algorithms to classify. The one to quantum variational training phase, in this phase, the authors used four steps to compute the hyperplane between training data and another to quantum variational classification phase to classify new datum to correct label. The second model is a SVM quantum kernel-based algorithm based on quantum kernel estimation. Maria Schuld et al. (41) proposed two- hybrid quantum techniques for classification problems. Schuld showed that quantum computing enhances classical ML algorithms like kernel methods. Quantum computing performs complex computations in Hilbert space more efficient. The authors focused on using feature maps and kernel methods in the quantum computing world.

Mitarai, K et al (42) presented a hybrid quantum-classical technique to perform different tasks like classification, regression, and clustering to can implement it on the small-scale quantum devices. Jessica Pointing introduced (43) in the Ph.D. thesis a novel quantum-classical algorithm to handing missing values in data is named "a Quantum Algorithm for Handling Missing Data". The main advantage of the algorithm calculates the probability distribution to handle missing data values.

Ruan, Y et al. (44) presented a quantum KNN algorithm based on the hamming distance matrix. QKNN is a good analog for the classical KNN algorithm, which avoids the defect of the simplified assumption of the task of classification QKNN outperforms Centroid and QNN on time performance and classification accuracy. Grant, E et al. (45) introduced a new hybrid classifier based on a hierarchical structure for quantum circuits for binary classification problems. Zhang, D. B et al (46) presented a quantum version of nonlinear regression rely on the hybrid quantum device. also, the authors proposed a new encoding method. Benedetti, M. (47) introduced a framework generative model named "data-driven quantum circuit learning" (DDQCL). DDQCL approach is unsupervised hybrid quantum- classical for the characterization of NISQ hardware for solving sampling problems such as bars and stripes (BAS) and random thermal

5data sets by using shallow quantum circuits.

| Research | Model | Class | Task | Application |
|---|---|---|---|---|
| Soumik et al (38). | Single-shot training | Hybrid | Classification | Cancer, Sonar, and Iris |
| Sanjay, et al. (39) | HQFSA | Hybrid | Feature selection | Breast Cancer, Iris, Wine, Vehicle, Glass, Sonar, and Ionosphere. |
| Dennis, et al. (22) | QA_SVM | Quantum | Classification and optimization | Real and synthetic data |
| Prayag et al. (27) | QIBC | In-spired | Classification | Text and image Corpora data set |
| Sergioli et al. (28) | HQC | In-spired | Classification | - |
| da Silva, A. et al (25) | QPF | Quantum | Classification | - |
| Ding et al. (29) | QiSVM | In-spired | Classification | - |
| Casaña-Eslava, R. (35) | Probabilistic Quantum Clustering (PQC) | In-spired | Clustering | Local densities & Two spirals |
| Yu, C.(36) | - | In-spired | Regression | - |
| Benedetti, M. (47) | DDQCL | Hybrid | Generative model | Synthetic data |
| Jessica Pointing (43) | - | Hybrid | Handling missing values | Heart data set |
| Maria Schuld et al. (41) | - | Hybrid | Classification | - |
| Lu, S et al (33) | MERA and TTN | Hybrid | Classification | MNIST digit and synthetic data set |
| Sergioli et al. (30) | QNMC | In-spired | Classification | Medical Data |
| Dang, Y et al. (31) | QKNN | In-spired | classification | Graz-01 and Caltech-10 datasets |
| Schuld, M et al (48) | Circuit-centric quantum classifier | Hybrid | classification | Cancer and MNIST |
| Bishwas, A. (34) | Binary clustering QSVM | In-spired | Clustering | Big Data |
| Zhang, D. B et al (46) | | Hybrid | Regression | |
| Ruan, Y et al (44) | QKNN | Hybrid | Classification | MNIST digit data set |
| Schuld, M et al (26) | Quantum linear regression | Quantum | Regression | - |
| Chen, H. et al (32) | QKNN | In-spired | Classification | - |
| Rebentrost, P et al. (24) | QSVM | Quantum | Classification big data | - |
| Lu, S et al (33) | Quantum decision tree | In-spired | Classification | - |
| Sagheer, A et al (37) | APM | In-spired | Classification | Breast cancer and synthetic data |
| Mitarai, K et al (42) | QCL | Hybrid | Classification & Clustering | - |

**TABLE 1** Summarizes quantum machine learning algorithms

| Algorithm | Classification Type | Based Idea | Q Data | Data Set | Advantages | Limitations |
|---|---|---|---|---|---|---|
| QIBC | Binary | Quantum detection theory | ✓ | image and text corpora data set | High precision, recall, and F-measure. Less computation time compared to SVM and KNN. | Need to many features to improve performance. Increase computation cost with large number copies. |
| HQC | Binary | distinguishability between quantum states | ✓ | Appendicitis and other data sets | Improve classification rate. | Don't work efficiently with many features vectors in large data set. |
| QNMC | Binary & Multi-class | quantum formalism | ✓ | Diabetes, Liver Cancer Ionosphere | Achieve higher accuracy than classical NMC | Don't achieve better performance with cancer data. |
| variational quantum classifier | Binary & Multi-class | Quantum Variational circuit | ✓ | Breast Cancer and wine data | Higher performance Classy binary & multi-class efficiently. | High cost |
| QSVM | Binary & Multi-class | Run on real quantum device | ✓ | Breast Cancer and wine data set | Better accuracy from classical SVM. | Difficult implementation |
| APM | Binary & Multi-class | Quantum Bit | x | Breast Cancer Wine Vintage synthetic data | Classify specific nonlinear problems with one neuron only. Take a few instances to train model and achieve higher accuracy. | Work on classical computer only. |

**TABLE 2**
A comparative table among QML classification from through classification type, based idea for algorithm, Q data refers to algorithm works on quantum data (✓) or not (x), data sets that algorithm use it for evaluation the performance, Advantages

## 3 | QUANTUM CLASSIFICATION SCHEME

Throughout this section, we present the concept of quantum classification, encoding methods, variational quantum circuits and Finally, we discuss the quantum classification scheme.

**a) Classification concept in quantum domain**

Classification is a popular task in a supervised learning domain that maps given inputs data (x) to discrete target output (y) through a function approximation f as follows y = f(x). The main purpose of classification is an accurate prediction model. Classification problems can be classified into two main problems: classification with two-classes is known as binary classification (for instance cancer diagnosis and spam detection). Classification with more than two-classes is known as multi-class classification (for instance image and digit classification). classification problem can be represented with classical ML domain as C = {$c_1, c_2, ..., c_n$} where C is target labels and a set of data in training phase with as $D_n$= {$(x_1, y_1), ..., (x_i, y_i), ..., (x_n, y_n)$} where xi is some of the features (n) on the properties of order of data point (i) and $y_i$ is the corresponding of that data point. In the case of binary classification $y_i \in \{c_1, c_2\}$ and xi $\in R^d$. in the case of multi-class classification $y_i \in \{c_1, ..., c_n\}$ where xi $\in R^d$ and d is real-valued attributes. To can describe classification problems with the QML domain, we should convert classical data to quantum data then we can be represented quantum data in training data as $D_n$ = {( $|\psi_1\rangle, y_1$ ), ... , ( $|\psi_i\rangle, y_i$ ),..., ( $|\psi_n\rangle, y_n$)} where $|\psi_i\rangle$ is the order (i) of the quantum state of $D_n$, $|\psi_i\rangle \in C^{2^d}$ and $y_i \in$ {c1, c2 }[33],(49) .in the case of binary classification, How to map data to quantum form? There is more than method (or technique) for encoding classical data to quantum data such as basis encoding and amplitude encoding for in detail (in next subsection b)- Encoding methods)

**b) Encoding methods**

There are a lot of encoding methods from classical data to quantum data in a Hilbert space. In other words, encoding data means loading classical data into quantum computer (quantum states). We outline two methods. for other methods, we refer the reader to (41) (50) (72).

**Basis Encoding** is the simplest method to encode data to quantum data. This method associates between n-bit classical input and the computational basis of n-qubit input. For example, (1100) classical input string is encoded to four qubits ( $|1100\rangle$) quantum states. In general, to encode data set by basis method, in other word, to represent data in computational basis states of qubits, we use the following equation.

$$|D\rangle = \frac{1}{\sqrt{M}} \sum_{m=1}^{M} |X^m\rangle \qquad (2)$$

Where D = $\{X^1, X^2, \ldots, X^M\}$ is classical data that is in form binary string , $X^m = \{b_1, b_2, \ldots b_N\}$, $b_i \in \{0,1\}$ , i ∈ { 1,2,....,N} and N is number of features.

**Amplitude Encoding** is the most used and popular method encoding in QML algorithms is amplitude encoding. The main idea of amplitude encoding based on association classical data with quantum state amplitudes. To encode classical data vector to quantum amplitudes. We should convert the classical vector to a normalized classical vector (28,49).

$$X = \begin{bmatrix} x_1 \\ x_2 \\ \vdots \\ x_{2^n} \end{bmatrix} \qquad (3)$$

Where X is normalized classical vector , x ∈ $C^{2^n}$, and is C complex numbers.
quantum state amplitudes can be encoded as follows:

$$|\psi_x\rangle = \sum_{i=1}^{2^n-1} x_i |i\rangle \qquad (4)$$

where $|\psi\rangle \in$ Hilbert space($\mathcal{H}$) and $\sum_i |xi|^2 = 1$

**c) Variational quantum circuit**

Variational quantum circuits also known as parameterized quantum circuits, most hybrid quantum-classical algorithms rely on these circuits (51). The key idea of the variational circuit is optimizing the parameters according to an objective function. variational quantum circuits consist of two phases (see Figure 1): the quantum phase and classical phase. The quantum phase includes state preparation, quantum circuit is the heart of the variational circuit that parameterizes input (X) based on numbers of parameters (θ) and measurement. The classical phase includes the output of the circuit, objective function and learning algorithm." quantum variational circuit" can be optimized by classical optimization algorithms as gradient descent, stochastic gradient descent and particle swarm optimization. some of the uses of variational quantum circuit such as machine learning (40-41), optimization (52), deep learning (53) and The variational circuit also uses in solving the complicated optimization problems.

**d) Classical proposed quantum classification scheme**

We propose a new a quantum classification scheme consists of six phases as collect datasets, pre-processing & encoding, training, validation (evaluate and optimize model), testing phase and decoding phase. (as shown in Figure 2-part A). Also, we show a classical classification scheme (as shown in Figure 2 -part B), we describe these phases as following:

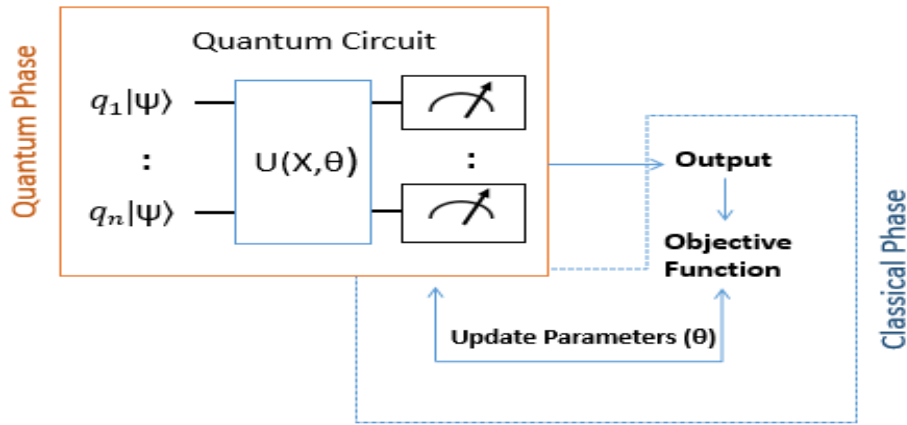

**Figure 1**. Illustration two phases of the variational circuit (parametrized quantum circuit) for Hybrid quantum-classical applications in a quantum machine learning (QML) and quantum neural networks (QNNS). Quantum phase that includes a quantum circuit that consists of state preparation, circuit, and measurement. another phase is a classical phase that includes output after measurement and consists of cost function and learning algorithms which updates parameters (θ).

1. collect datasets

According to the classifier, we can determine whether it works on quantum or classical data. Some of the quantum-inspired ML works on classical data for instance (37). and other works on quantum data (27). Also, Hybrid quantum-classical classifiers work on quantum data as (33).

2. Pre-processing & Encoding

The preprocessing phase may be applied by classical or quantum techniques. For example, for handling missing values, we can use classical or quantum techniques. Then, reduce number of the features to are compatible with available limited numbers of qubits on quantum device and split data to training, validation and test set. Then, apply the encoding for mapping data from classical data (X) to quantum data ( $|\psi_x\rangle$) in order to prepare data as quantum states. Next, applying a quantum classifier.

3. Training

Training phase varies according to quantum classifier type. For example, in the case of quantum-inspired classifier, we train model on classical machine. In the case of quantum kernel classifier, we estimate kernel on a quantum device) for instance, IBM Q computer or quantum simulator) .

4. Validation

We evaluate and update parameters to decrease cost function and improve performance of model.

5. Testing

We test model performance by confusion matrix and calculate test time.

6. Decoding (or readout)

To calculate output data and translate into classical form (i.e. C = $\{-1,1\}$ ).

## 4 | DISCUSSION

Quantum algorithms are quantum models when running a real quantum computer on but don't run their classical counterparts on such as Shor's algorithm (8), Grover's algorithm (9), addition on a quantum computer (54). for more details about these algorithms, see (13), (25), (55).

Quantum machine learning (QML) has become a new growth research field and has appeared in many applications and countries. the progress and success of quantum computing have observed significantly So, we need to apply the advantages and properties of QM with the machine learning branch.

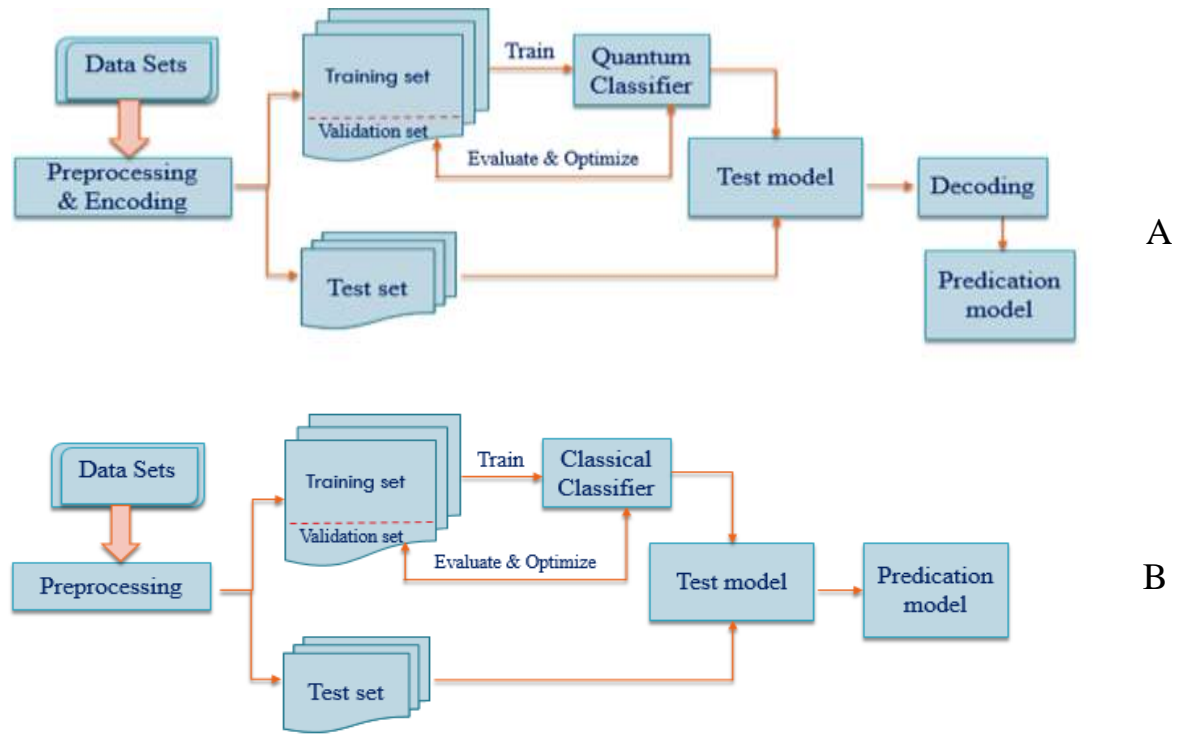

**Figure 2.** proposed quantum scheme for classification and prediction pipeline.
Above part (A): the quantum classification and prediction scheme consist of preprocessing, encoding, train, test, and decoding steps. The quantum classifier requires two main steps encoding and decoding (readout) data. First, the step of preprocessing & encoding in the quantum model may be applied by classical or quantum techniques. Many quantum classifiers work on data and others work on classical often in quantum-inspired classifiers. In the case of quantum inspired classifiers that work on quantum data, we should map classical data (X) to quantum data ( $|\psi_x\rangle$) .thirdly, training phase, this step varies by quantum classifier type. For example, in the case of quantum-inspired, we train a model on the classical machine. In case of quantum kernel classifier, we kernel on a quantum device. Last, Decoding (measurement) to convert output labels data to classical. Below part (A): classical classification and prediction pipeline consists of preprocessing, Training & validation, testing and predication phase.

Due to the success of Shor's algorithm for solving factorization problems with exponential speedup, Grover's algorithm search problem in unstructured data with a quadratic speedup, and HHL algorithm (56) to solve linear systems equations. Quadratic Unconstrained Binary Optimization problem is NP-hard problems that are hard to the classical hardware and algorithms solve it. QUBO problem is solved by quantum annealing(57) (see table 3) .

It's time to discuss the survey questions: Why do we proceed towards Quantum Machine Learning (QML)? To solve challenges and difficult tasks that still exist in classical machine learning. With increasing size of data, learning phase training time and memory consumption to increase during the learning phase. to achieve low-cost learning with higher accuracy and performance is a challenging task with classical algorithms. It's hard to estimate kernel functions in higher dimensions (58-60). it hard to determine the eigenvector (60) and solve complex optimization problems.

| Problem | Solved By | Speed up |
|---|---|---|
| Factorization | Shor's algorithm (8) | Exponential |
| Searching in unstructured data | Grover's algorithm (9) | Quadratic |
| Linear systems equations | HHL algorithm (56) | Exponential |
| Quadratic Unconstrained Binary Optimization (QUBO) | D-Wave quantum computer (70) | - |
| | IBM's Neurosynaptic hardware (71) | |

**TABLE 3**  Problems solved by quantum computing.

According to Dunjko, V et al. (15) Quantum ML can improve in run-time and efficiency. D-wave introduced Quadrant algorithms based on Central Processing Unit (CPU), Graphics Processing Unit (GPU) and Quantum annealing computer. Quadrant ML algorithms can achieve more performance with low-cost training data with large data over classical deep learning and traditional algorithms (61) (62).

Secondly, how does the concept of quantum computing enhance classical machine learning? Quantum computing can boost traditional algorithms by two different methods: The first method, implement classical algorithms on quantum computers or simulators. This method needs to encode classical data into quantum data. Another method, build QML algorithms depend on Quantum algorithms or quantum subroutines (63) such as Amplitude Amplification, Grover's algorithm, Quantum matrix-inversion, the quantum phase estimation, Variational quantum (48), quantum annealing, Sampling (47).

Quantum Fourier Transform (QFT) (64) is the heart of other quantum algorithms such as Quantum Phase Estimation algorithm and Shor's algorithm. Quantum Fourier Transform is a quantum version of inverse discrete Fourier Transform. QFT achieves exponential speed-up. The idea of QFT depends on mapping amplitudes of the current quantum state (R) to the next amplitudes of the quantum state (Q). $x_i$
Where $|R\rangle = \sum_{i=1}^{N-1} x_i |i\rangle$ , $|Q\rangle = \sum_{j=1}^{N-1} y_j |j\rangle$ with $x_i$, $y_j$ are complex numbers and mapping equation as following [60,75]:

$$Q_j = \frac{1}{\sqrt{N}} \sum_{n=1}^{N-1} X_n \, e^{2\pi i \frac{jk}{N}} \qquad (5)$$

Where j= 1, 2,….,N-1.

Quantum Phase Estimation (QPE) (65, 66) is one of the most important subroutines in many quantum algorithms [8,63] and quantum machine learning algorithms. QPE algorithm is based on Quantum Fourier Transform (QFT) subroutine. The goal of Quantum Phase Estimation (QPE) algorithm is to find the eigenvalue ($e^{2\pi i \, \theta}$) of the unitary matrix (U) or find the optimal approximation of the phase ($\theta$) as the following equation.

$$U|\varphi\rangle = e^{2\pi i \, \theta} |\varphi\rangle \qquad (6)$$

Where $0 \leq \theta < 1$ and $|\varphi\rangle$ is eigenvector.

Amplitude Amplification (AA) (67) (also called Quantum Interference) is one of the fundamental subroutines and is the key idea for the famous Grover's algorithm, and Grover's algorithm is a subroutine for many different algorithms. The key objective of Amplitude Amplification (AA) is boosting the solution probability of the amplitude (P) from arbitrary state to target state overall iterations (68, 69). To make amplify the successful probability of the amplitude by using the following formula:

$$1 - \left(\frac{m^2}{3}(P)\right)(m^2(P)) \qquad (7)$$

Where m is the number of iterations and P is success probability.

From table 4 that summarizes quantum subroutines and applications of each subroutine. We note that quantum Fourier transform (QFT) and variational quantum circuit is the most common and used subroutines in most algorithms and applications.

| Subroutine | Applications | Refs. |
|---|---|---|
| Amplitude amplification/Grover's algorithm | Quantum counting and searching | (9,67) |
| sampling | Quantum deep learning &" boson sampling" | (47) |
| Quantum phase estimation | Quantum counting | (8, 55) |
| Quantum Matrix- inversion/HHL Algorithms | QSVM, kernel least squares and machine learning | (55,56) |
| Quantum Fourier Transform | Shor's algorithm, Cryptography, Information processing, Communication, Discrete logarithm and Phase estimation | (13,54,64) |
| Variational quantum circuit | Optimization, classification | (33,48,40,51,53) |
| Quantum Annealing | Eigen state solver Machine learning Optimization Nurse Scheduling Healthcare | (22), (72, 73), (74) |

**TABLE 4** show quantum subroutines that enhanced classical ML techniques, algorithms applied based on it. To build QML algorithms and some of subroutines applications.

## 5 | CHALLENGES AND FUTURE DIRECTIONS

In this section, we outline and present challenges and future directions in QML as small-scale quantum computer, Limited quantum bits, encoding methods and develop a new QML techniques.

1. *Small-scale quantum computer*

the big and key challenge is to build quantum computers with a large number of qubits to implement, test QML algorithms and work with large different data in the near future. As shown in Figure 3, numbers of qubits achieved by different technology companies (75) such as Rigetti, IBM, Q-Wave, Xanadu, Google, and Microsoft. Up to now, the quantum computer is developed in small-scale and this restricts us use a limited amount of data (76). So, researchers develop algorithms work on available small-scale and "noisy intermediate-scale quantum" (NISQ) quantum hardware compatible with number of qubits hardware(77).limited qubits number using a less number of features that lead to losing a lot of important data and with limited qubits can't apply big data processing on quantum devices (78, 79) .

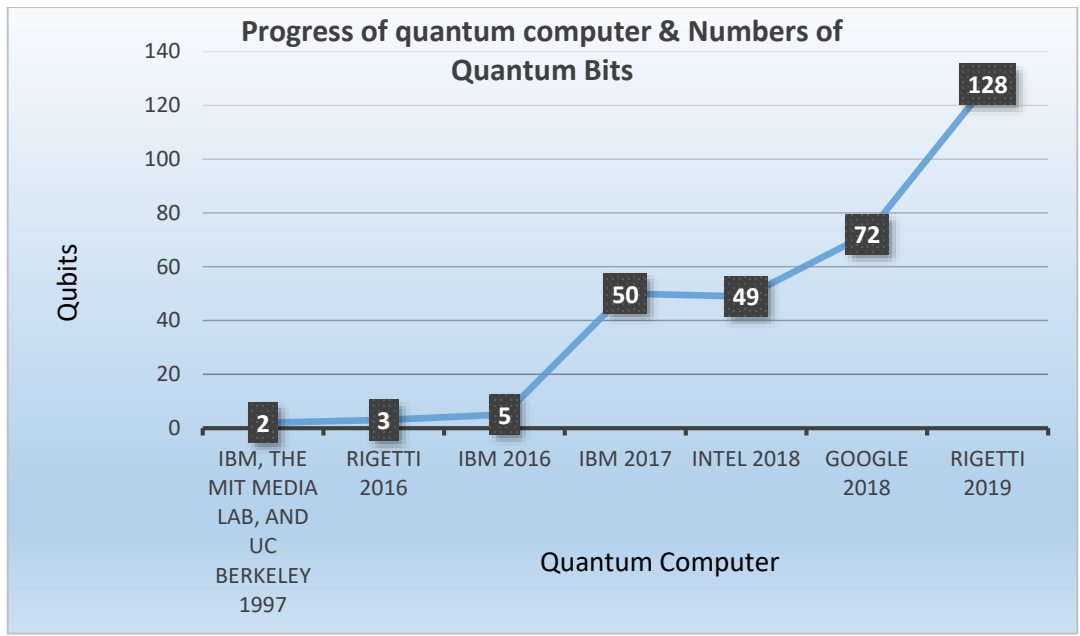

**Figure 3.** A chart shows the progress of quantum computer with numbers Quantum Bits: Source (75), Numbers of qubits achieved by different companies. The first quantum computer with 2-qubits build by IBM, the MIT Media Lab, and UC Berkeley in 1997.IBM achieved in 2017, 50 qubits.20 years to increase 48 qubits .in 2019, Rigetti achieved 128 qubits.in 2 years 78 qubits increased. If numbers of qubits increase with a high error rate don't make a powerful quantum processor (80). Powerful quantum processer with a low error rate will increase exponentially by increasing numbers of quantum bits.

## 2. Encoding methods

Encoding data to quantum states is one of QML challenges, this process takes high consumption time and power for mapping from classical data into quantum data (74) (i.e. image and big data). So, design new techniques for encoding data is an interesting future direction for research. In 2020, LaRose et al (81).presented a binary quantum classifier in order to robust with noise quantum states based on the selection of the best encoding method to load data into the quantum system. Furthermore, the authors discussed different methods for encoding with quantum binary classifier. also, the authors applied several encoding data on the same data and showed that encoding data techniques can be improved the model accuracy. In 2020, the authors (74) introduced a new method for encoding large amount of image data into a few number of qubits after compression to reduce of image based on "quantum annealing computer. As well, the authors trained "restricted boltzmann machine" to classify images data.

## 3. A new QML techniques

In near future, we think that researchers will develop new theoretical and applied ML algorithms that are compatible with available quantum hardware. using quantum information theory or quantum subroutines for solving ML problems and improve performance. As well, create versions for current algorithms in several fields. For example, quantum neural networks (QNNs), quantum deep Learning, quantum-inspired ML, and quantum-enhanced ML(82).in addition to, develop new classification techniques based on quantum "variational circuit" and working on reducing the depth of quantum circuit. And also, implementation of classical techniques on real quantum machines or simulators. using quantum annealing and adiabatic computation with classical ML to develop new QML paradigms and solve complicated optimization problems related to ML problems.

## 6 | APPLICATIONS

Quantum Machine Learning (QML) techniques are more effective in many real-world applications comparable to traditional machine learning in speedup and accuracy as big data classification [24], forecasting series, spam detection, image compression, medical domain (30,43,37) (i.e., cervical cancer detection (83), electronic calculations (84) ,decision games (85), natural language processing (NLP), recommendation systems (86, 87) , speech recognition (88), image classification (27,44) , and electrocardiogram signals classification (89) .Applications related to hybrid quantum-classical approach such as scheduling problems (90), and classification task (40-41).

## 7 | CONCLUSION

This survey paper reviewed the most recent articles for quantum machine learning techniques for various problems that improve and handle different problems more effectively and accurately from traditional techniques in conventional computing. And also, we compared between quantum-inspired and hybrid- quantum-classical algorithms. as mentioned in section 2. (literature review), QML algorithms outperformed classical ML in performance and speed up. In addition, we presented the quantum scheme for classification problems or in general in supervised learning then we outline two methods for mapping data. We discussed methods of quantum to enhance machine learning such as quantum subroutines. We presented applications of QML in different fields and future vision, to open new avenues for use of quantum machine learning in the research field. Our future work, we will apply quantum-inspired and hybrid quantum-classical algorithm on real-world data to compare its performance.

**ACKNOWLEDGEMENTS**
We are grateful to prof. Giuseppe Sergioli for helpful discussion.